\documentclass[osajnl,twocolumn,showpacs,superscriptaddress,10pt]{revtex4-1}

\usepackage{color}
\usepackage{amsfonts,amsmath,graphicx} 
\usepackage{epstopdf}%

\newcommand {\be} {\begin{equation}} 
\newcommand {\ba}{\begin{eqnarray}} 
\newcommand {\ee} {\end{equation}} 
\newcommand{\ea} {\end{eqnarray}}

\renewcommand{\Im}{{\rm Im\,}} 
\renewcommand{\Re}{{\rm Re\,}}
\renewcommand{\epsilon}{\varepsilon}

\preprint{MKPH-T-yy-xx}

\begin{document}

\title{
Novel Properties of Twisted-Light Absorption}

\author{Andrei Afanasev}

\affiliation{Department of Physics,
The George Washington University, Washington, DC 20052, USA}

\author{Carl E.\ Carlson}

\affiliation{Department of Physics, The College of William and Mary in Virginia, Williamsburg, VA 23187, USA}

\author{Asmita Mukherjee}

\affiliation{Department of Physics, Indian Institute of Technology Bombay, Powai, 
Mumbai 400076, India}



\begin{abstract}

We discuss novel features of twisted-light absorption both by hydrogen-like atoms and by micro-particles.
First, we extend the treatment of atomic photoexcitation by twisted photons to include atomic recoil, derive generalized quantum selection rules and consider phenomena of forbidden atomic transitions. Second, using the same electromagnetic potential for the twisted light beams, we analyze the radiation pressure from these beams on micro-sized particles, and verify that while the Poynting vector can in some circumstances point back toward the source, a complete analysis nonetheless gives a repulsive radiation pressure. 

\end{abstract}

\maketitle

\section{Introduction}	\label{sec:intro}

In 1992 Allen and collaborators~\cite{Allen:1992zz} pointed out that  Laguerre-Gaussian laser modes may carry large projections of angular momentum on the direction of beam propagation.
It became a very active field of studies, with many applications of optical vortices in optical tweezers, encoding of information, and microscopy. Dedicated recent reviews can be found in Ref. \cite{Yao11,TorresTornerBook}.  A term``twisted photon" is used to describe such beams at a quantum level~\cite{molina2007nature}. Optical vortices were successfully demonstrated in a broad range of photon energies, from radio-waves~\cite{2012NJPh...14c3001T} to X-rays~\cite{2013NatPh...9..549H}. It was also suggested to use Compton backscattering \cite{Jentschura:2010ap,Jentschura:2011ih} to obtain high-energy photons that may be relevant for nuclear and particle physics research.

To analyze the unique properties of the twisted photons, we considered their absorption by an atom. This problem was previously addressed by several authors~\cite{Picon10,picon2010njp,Davis13,Jauregui:2004}. Most of the above papers considered an atom placed in the center of an optical vortex. Atomic ionization in a more general case was addressed in Ref.~\cite{2013JPhB...46t5002M}.

In our recent paper on photoexcitation of atoms \cite{Afanasev:2013kaa} we made a (well-justified) approximation that neglected atomic recoil at the cost of apparent non-conservation of total angular momentum. Here we derive quantum selection rules that describe atomic transitions caused by twisted photons, with effects of atomic recoil included.

In addition, this paper focuses on two novel phenomena that twisted photons can produce.  In both cases we look at the low end of the angular momentum projected along the propagation direction $m_\gamma$ possibilities, namely $m_\gamma = 0$ and $m_\gamma = -\Lambda$, where $\Lambda = \pm 1$ is the usual plane wave helicity of the photon.  These possibilities cannot occur for single plane wave photons and are in the twisted photon realm.    Herein after we refer to $m_\gamma$ as an ``total helicity.'' 

Other recent publications considered forces on an atom due to a plane wave light pulse~\cite{hinds2009} and discussed the issue of azimuthal linear momentum of an optical vortex~\cite{speirits2013}.

Atomic transitions induced by $m_\gamma = 0$ photons can produce final states with quantum numbers forbidden to plane wave photon transitions.  For example, one can photoexcite ground state hydrogen by electric dipole transitions into $P$-state orbitals with magnetic quantum number $m_f$ equal zero (using the incoming wavefront's propagation direction as the quantization axis).  Plane wave photons produce $m_f = \Lambda = \pm 1$.  As a practical matter, there are factors associated with the way the twisted photon is composed that can suppress the rate, even though the electric dipole transition itself is leading order.  However, depending on circumstances, the rate could be large enough to be observable.  Details, along with background material on twisted photons and the discussion of atomic recoil effects, are given in Sec.~\ref{sec:atomicX}.

An examination of the Poynting vector for twisted photons shows regions, notably along the beam axis,  where the component along the beam direction is directed back to the source.  In Sec.~\ref{sec:tractor}, we consider in detail the radiation pressure on small particles that absorb such twisted photons, and discuss whether it leads to tractor effects.

Some concluding remarks are offered in Sec.~\ref{sec:disc}.


\section{Atomic excitation by twisted photons}			\label{sec:atomicX}



\subsection{Atomic states}


To establish notation, let the electron coordinate be $\vec r_1$ and the proton or nucleus coordinate be $\vec r_2$.  The relative and center of mass (CM) coordinates are
\begin{align}
\vec r &= \vec r_1 - \vec r_2 	\,,	\nonumber\\
\vec R &= \frac{1}{M} \left(m_1 \vec r_1 + m_2 \vec r_2 \right)		\,,
\end{align}
for $M= m_1 +m_2 = m_e + m_p$.  Likewise,
\begin{align}
\vec r_1 = \vec R + \frac{m_2}{M} \vec r	\,,	\nonumber\\
\vec r_2 = \vec R - \frac{m_1}{M} \vec r	\,.
\end{align}

Nonrelativistically, since the atomic binding potential depends only on $\vec r$, the wave function for the atomic state naturally separates into a product
\begin{align}
\Psi(\vec r, \vec R) = \psi(\vec r) \Psi(\vec R)	\,.
\end{align}


\subsection{Twisted photon expansion}


To define the twisted photons, we follow~\cite{Jentschura:2010ap,Jentschura:2011ih}, whose states may be viewed as extensions of the nondiffractive Bessel modes described in~\cite{Durnin:1987,Durnin:1987zz};  see also~\cite{Jauregui:2004}.  More detail is given in~\cite{Afanasev:2013kaa}; we give a short and hopefully sufficient summary here.   

A twisted photon state moving in the $z$-direction, with total helicity $m_\gamma$ and with symmetry axis passing through the origin, can be given as a superposition of plane waves and in Hilbert space can be written as, 
\begin{align}
\label{eq:twisteddefinition}
| \kappa m_\gamma k_z \Lambda \rangle 
&= \sqrt{\frac{\kappa}{2\pi}} \  \int \frac{d\phi_k}{2\pi} (-i)^{m_\gamma} e^{im_\gamma\phi_k}  \,
	|\vec k, \Lambda\rangle		\,.
\end{align}
The component states on the right are plane wave states, all with the same longitudinal momentum $k_z$, the same transverse momentum magnitude $\kappa = |\vec k_\perp|$, and same plane wave helicity $\Lambda$ (in the directions $\vec k$).  Angle  $\phi_k$  is the azimuthal angle of vector $\vec k$, and with the phasing shown, $m_\gamma$ is the total angular momentum in the $z$ direction, with the possibility that $m_\gamma \gg 1$.  We also define a pitch angle   $\theta_k = \arctan (\kappa/k_z)$, and $\omega = | \vec k |$.

The electromagnetic potential of the twisted photon in coordinate space, at the location of the electron is 
\begin{align}
\label{eq:twistedwave}
\mathcal A^\mu_{\kappa m_\gamma k_z \Lambda}(t,\vec r_1)
&= \sqrt{\frac{\kappa}{2\pi}} \, e^{-i\omega t} \nonumber\\
&\times	\int \frac{d\phi_k}{2\pi} (-i)^{m_\gamma} e^{im_\gamma\phi_k}  \,
	\epsilon^\mu_{\vec k,\Lambda} e^{i \vec k {\cdot} \vec r_1}	.
\end{align}
The polarization vectors are~\cite{Jentschura:2010ap,Jentschura:2011ih,Afanasev:2013kaa}
\be
\label{eq:epsilonexpand}
\epsilon^\mu_{\vec k \Lambda} \!\! = \!
	e^{-i\Lambda\phi_k} \! \cos^2\frac{\theta_k}{2} \eta^\mu_\Lambda
	+ e^{i\Lambda\phi_k} \! \sin^2\frac{\theta_k}{2} \eta^\mu_{-\Lambda}
	+ \frac{\Lambda}{\sqrt{2}} \sin\theta_k  \eta^\mu_0
\ee
with $4$-dimensional unit vectors,
\be
\eta^\mu_{\pm 1} = \frac{1}{\sqrt{2}}  \left( 0,\mp 1,-i,0 \right)	\,,
\quad \eta^\mu_0 =  \left( 0,0,0,1 \right)	\,.
\ee

\begin{figure}[h]
\begin{center}

\includegraphics[width = 67 mm]{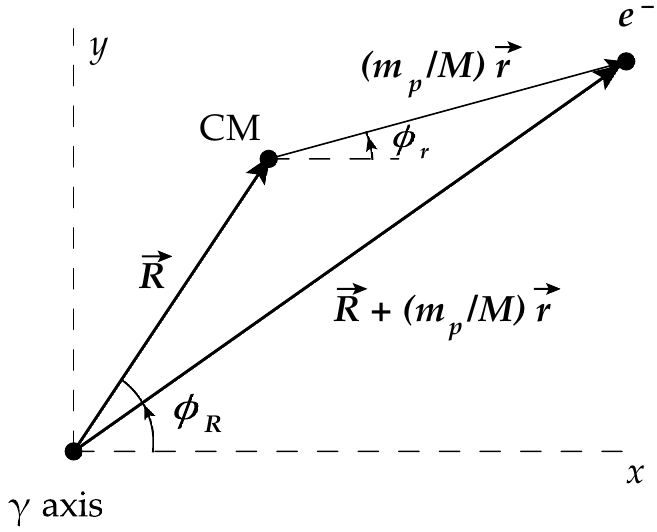}

\caption{Relative position of the atomic electron and the photon axis, as projected onto the $x$-$y$ plane, with the axis of the twisted photon passing through the origin and with the atomic center-of-mass (CM) also indicated. The proton is located at coordinate $-(m_e/M)\vec r$ relative to the CM.}
\label{fig:impact}
\end{center}
\end{figure}

It will be useful to expand the potential using cylindrical coordinates and the Jacobi-Anger formula, wherein
\begin{align}
e^{i \vec k_\perp {\cdot} \vec r_{1\perp}} &= e^{i \vec k_\perp {\cdot} \vec R_\perp} \, 
	e^{i \vec k_\perp {\cdot} ({m_p}/{M}) \vec r_\perp}
			\nonumber\\
&= \sum_{n_1,n_2 = -\infty}^\infty  i^{n_1+n_2} \, e^{i n_1(\phi_R - \phi_k)} \,
	e^{i n_2(\phi_r - \phi_k)}
			\nonumber\\
&\hskip 5 em \times	J_{n_1}(\kappa R_\perp) 
	J_{n_2} \big( \frac{m_p}{M} \kappa r_\perp \big)	\,,
\end{align}
where $R_\perp = | \vec R_\perp |$, $r_\perp = | \vec r_\perp |$, the $J_n$ are Bessel functions, and the azimuthal angles $\phi_R$ and $\phi_r$ are indicated in Fig.~\ref{fig:impact}.  After some manipulation, the photon potential becomes
\begin{align}
\label{eq:pot}
&\mathcal A^\mu_{\kappa m k_z \Lambda}(t,\vec r_1) =
	\sqrt{\frac{\kappa}{2\pi}} \, \frac{\Lambda}{i} \, e^{i(k_z z_1 - \omega t)}
			\nonumber\\
&\quad	\times	\sum_{n_1}	e^{i n_1 \phi_R} \, e^{i(m_\gamma - n_1) \phi_r}
				J_{n_1}(\kappa R_\perp)		\nonumber\\
&\quad	\times	
		\bigg\{	\cos^2 \frac{\theta_k}{2} \, \eta^\mu_\Lambda \, e^{-i\Lambda \phi_r} \,
		J_{m_\gamma - \Lambda - n_1}\big( \frac{m_p}{M} \kappa r_\perp \big)
			\nonumber\\
&\qquad	+ \frac{i}{\sqrt{2}} \sin\theta_k \, \eta^\mu_0  \,
		J_{m_\gamma-n_1}\big( \frac{m_p}{M} \kappa r_\perp \big)
			\nonumber\\
&\qquad  -	\sin^2 \frac{\theta_k}{2} \, \eta^\mu_{-\Lambda} \, 
		e^{i\Lambda \phi_r} \,
		J_{m_\gamma + \Lambda - n_1}\big( \frac{m_p}{M} \kappa r_\perp \big)
		\bigg\}	\,.
\end{align}


\subsection{Photoabsorption amplitude}


We shall be interested in calculating the transition probabilities between the ground state and excited states of the hydrogen atom caused by photon absorption.    The interaction Hamiltonian will be~\cite{schiff},
\begin{align}
\label{eq:hamiltonian}
H_1 = \sum_{j=1}^2 - \frac{e_j}{m_j} \vec A \cdot \vec p_j \,.
\end{align}
The matrix element for photoabsorbtion is
\begin{align}
&	i\mathcal M = \langle F,f | H_1(t=0) 
	| I,i; \gamma_{\kappa m_\gamma k_z \Lambda} \rangle 
		= \sum_j -\frac{e_j}{m_j}
			\nonumber\\
&\times	\int d^3R \ d^3r \ \Psi^*_F(\vec R) \psi^*_f(\vec r)
	\vec A(0,\vec r_j) {\cdot} \vec p_j   \Psi_I(\vec R) \psi_i(\vec r)	.
\end{align}
We have written the wave functions in factorized form, where $F$ and $f$ refer to the final state CM and relative wave functions, with $f$ later further specified by the quantum numbers $n_f, l_f$ and $m_f$, and analogously $I$ and $i$ refer to initial state CM and relative wave functions, with $i$ being short notation for $n_i, l_i$ and $m_i$. 

With two different momenta, proton and electron, and with each momentum becoming a derivative that acts on both CM and relative coordinates, one has four contributions to the above amplitude.  We will explicitly work out the formulas for the electron case, which should be by far the largest contributions, and consider the derivatives on the CM and relative momenta serially.  For the electron momentum,
\begin{align}
\vec p_1 = \frac{m_e}{M} \vec P + \vec p = -i \frac{m_e}{M} \vec\nabla_R - i \vec\nabla_r
\end{align}
Because of the electron mass is so small compared to the total mass, the second term should be by far the larger, and we consider it first.  

Also, we will take the initial wave function of the atomic electron to be in the ground state, 
$\psi_i(\vec r) = \psi_{n_i l_i m_i}(\vec r) \to \psi_{100}(\vec r) = R_{10}(r) Y_{00}(\theta_r,\phi_r)$.


\subsection{Relative momentum contribution}


For this subsection we keep just the electron's $-i\vec\nabla_r$ contribution, and using 
\begin{align}
	-i \hat\eta_\lambda {\cdot} \vec\nabla_r \, R_{10}(r) Y_{00} 
	= -i \sqrt{\frac{1}{3}} Y_{1\lambda}(\theta_r,0) e^{i\lambda \phi_r} R'_{10}(r)
\end{align}
for $\lambda = \pm 1, 0$, we obtain after doing the $d\phi_r$ integration,
\begin{align}
&	i\mathcal M_{e,rel} = - \frac{e_1 \Lambda}{m_e a_0} \sqrt{\frac{2\pi\kappa}{3}}	\,
				G_{FI}
				\nonumber\\
&\times	\left\{ \cos^2 \frac{\theta_k}{2} g_{f \Lambda} 
	+ \frac{i}{\sqrt{2}} \sin\theta_k \, g_{f 0}
	- \sin^2 \frac{\theta_k}{2} g_{f,-\Lambda}	\right\}	.
\end{align}

Here,
\begin{align}
\label{eq:GFI}
G_{FI} &= \int d^3R \ \Psi_F^*(\vec R) \Psi_I(\vec R) 
				\nonumber\\
&\times	e^{i(m_\gamma-m_f)\phi_R}
	J_{m_\gamma-m_f}(\kappa R_\perp)		e^{ik_z Z}	.
\end{align}
Further, the relative coordinate integrals give the atomic factors $g_{f\lambda} = g_{n_f l_f m_f \lambda}$ as in~\cite{Afanasev:2013kaa}, up to some near unity factors $(m_p/M)$,
\begin{align}
&g_{f\lambda} = - a_0 \int_0^\infty r^2 dr \, R_{n_f l_f}(r)
	R'_{10}(r) \int_{-1}^1 d(\cos\theta_r)
			\nonumber\\
&\, \times J_{m_f-\lambda} \big( \frac{m_p}{M}\kappa r_\perp \big) 
	Y_{l_f m_f}(\theta_r,0)  Y_{1\lambda}(\theta_r,0)  e^{i(m_p/M) k_z z}	.
\end{align}

We consider two special cases for the CM wave function.
If the CM wave function is well localized, we may center it in the $Z=0$ plane at a definite distance and azimuth angle from the photon axis, and approximate the unit normalized state by a function sharply peaked around $R=b$, $\cos\theta_R = 0$, and $\phi_R = \phi_b \pm \pi$.

If in addition we neglect recoil and use the same wave function for the final state, we obtain,
\begin{align}
\label{eq:localized}
&	i\mathcal M_{e,rel} = - \frac{e_1 \Lambda}{m_e a_0} \sqrt{\frac{2\pi\kappa}{3}}
		e^{i(m_\gamma-m_f)\phi_b}
		J_{m_f - m_\gamma}(\kappa b)
				\nonumber\\
&\times	\bigg\{ \cos^2 \frac{\theta_k}{2} g_{n_f l_f m_f \Lambda} 
	+ \frac{i}{\sqrt{2}} \sin\theta_k \, g_{n_f l_f m_f 0}
				\nonumber\\
&\hskip 1.3 em	- \sin^2 \frac{\theta_k}{2} g_{n_f l_f m_f,-\Lambda}	\bigg\}	,
\end{align}
in agreement with~\cite{Afanasev:2013kaa}.

If the initial and final CM wave functions have definite eigenvalues of the angular momentum operator $L_Z$~\cite{Jauregui:2004}, then we can consider transitions between CM states,
\begin{align}
\Psi_I(\vec R) &=  \frac{e^{im_R \phi_R}}{\sqrt{2\pi}} \mathcal Y_I(R,\cos\theta_R)	,	\nonumber\\
\Psi_F(\vec R) &= \frac{e^{im'_R \phi_R}}{\sqrt{2\pi}} \mathcal Y_F(R,\cos\theta_R)	.
\end{align}
Substituting these wave functions into Eq.~(\ref{eq:GFI}) gives,  
\begin{align}
\label{eq:amplitude}
&i\mathcal M_{e,rel} =  \frac{ e \Lambda}{m_e a_0} \sqrt{\frac{2\pi\kappa}{3}} \,
	\delta_{m'_R - m_R, m_\gamma - m_f} \, \tilde G_{FI}	
				\nonumber\\
&\times	\left\{ \cos^2 \frac{\theta_k}{2} g_{f \Lambda} 
	+ \frac{i}{\sqrt{2}} \sin\theta_k \, g_{f 0}
	- \sin^2 \frac{\theta_k}{2} g_{f,-\Lambda}	\right\}	\,,
\end{align}
with Kronecker deltas appearing after integration over the azimuthal CM angle $\phi_R$, and leaving the remaining integrals in the form,
\begin{align}
\label{eq:twiddle}
\tilde G_{FI} &= \int_0^\infty R^2 dR \int_{-1}^1 d(\cos\theta_R)
			\nonumber\\
&\ \ \times	\mathcal Y_F^*(R,\cos\theta_R)	\mathcal Y_I(R,\cos\theta_R)
	J_{m'_R - m_R}(\kappa R_\perp)  e^{ik_z Z}	\,.
\end{align}

Even more particularly, if the $\mathcal Y$ wave functions are well localized in radius and angle,  $\tilde G_{FI} = J_{m_\gamma - m_f}(\kappa b)$, and the amplitude~(\ref{eq:amplitude}) has the same magnitude as~(\ref{eq:localized}).

Eq.~(\ref{eq:amplitude})  contains a $\delta$-function showing conservation of angular momentum along the $z$-direction.  Whatever part of the photon's total helicity that does not go into exciting the atomic wave function goes into making the atom as a whole revolve about the photon's symmetry axis.

If the initial CM wave function is centered in the vicinity of some definite azimuthal point, as in Fig.~\ref{fig:impact}, then the initial wave function could be expanded as a sum of $L_Z$ eigenfunctions over a range of $m_R$.  We can work, as we do here, with the individual terms in such a sum, noting that if the target atom is not revolving about the origin, then the expectation value $\langle m_R \rangle = 0$.  In general,
\be
\langle m'_R \rangle = \langle m_R \rangle +m_\gamma - m_f		\,.
\ee

J\'auregui has obtained similar results~\cite{Jauregui:2004}.  The methods used here lead to results that appear qualitatively simpler and more compact, and without lingering summations in the final results, for the CM wave functions displayed.  No dipole approximation has been used, so that excitation of highly excited atomic states can be accurately calculated.


\subsection{CM derivative contribution}


Continuing to work out the case where the electron absorbs the photon, we now consider the $\vec\nabla_R$ term in 
$p_1 = -i(m_e/M)\vec\nabla_R - i \vec\nabla_r$.  We have
\begin{align}
&i\mathcal M_{e,R} = \frac{ie_1}{M} 
	\int d^3r \, R_{n_f l_f}(r) Y_{l_f m_f}(\theta_r,0)
	e^{-im_f \phi_r} 
			\nonumber\\
&\hskip 1 em \times	R_{10}(r) Y_{00}	\int d^3R \, \Psi^*_F(\vec R) 
	\vec A(0,\vec r_1) {\cdot} \vec\nabla_R 	\Psi_I(\vec R)	\,.
\end{align}

We again expand the CM wave functions as, for the initial state, 
$\Psi_I(\vec R) = e^{im_R \phi_R} \mathcal Y_I(R_\perp,Z)$, and let
\begin{align}
\hat\eta_\lambda {\cdot} \vec\nabla_R \Psi_I(\vec R)
	= e^{i(\lambda+m_R) \phi_R} \, \partial_{R\lambda}\mathcal Y_I(R_\perp,Z)	\,,
\end{align}
where
\begin{align}
\partial_{R\lambda} \mathcal Y_i= \left\{
	\begin{array}{ll}
	\frac{1}{\sqrt{2}}\left( -\Lambda \frac{\partial}{\partial R_\perp} 
					+ \frac{m_R}{R_\perp} \right) \mathcal Y_I
					\,, & \lambda=\Lambda = \pm 1	  \,,	\\[1.2 ex]
	\frac{\partial}{\partial Z} \mathcal Y_I \,,
		& \lambda = 0	\,.
	\end{array}
				\right.
\end{align}
After performing the $d\phi_R$ and $d\phi_r$ integrals, one obtains
\begin{align}
\label{eq:amplitudeCM}
&i\mathcal M_{e,R} =  \frac{2\pi e \Lambda}{M a_0} \sqrt{\frac{\kappa}{2}} \,
	\delta_{m'_R - m_R, m_\gamma - m_f} \, g_{fi}	
				\nonumber\\
&\times	\!  \left\{ \cos^2 \frac{\theta_k}{2} G_{FI,\Lambda} 
	+ \frac{i}{\sqrt{2}} \sin\theta_k \, G_{FI,0}
	- \sin^2 \frac{\theta_k}{2} G_{FI,-\Lambda}		\!  \right\}	\!\!.
\end{align}
This time,
\begin{align}
&g_{fi} = \int_0^\infty  r^2 dr\  R_{n_f l_f}(r)  R_{10}(r)  
	\int_{-1}^1 	d(\cos\theta_r)		\nonumber\\
&\hskip 2.5 em	 \times J_{m_f} \big( \frac{m_p}{M}\kappa r_\perp \big) 
	Y_{l_f m_f}(\theta_r,0)   e^{i(m_p/M) k_z z}	\,,
\end{align}
and
\begin{align}
\label{eq:GFI3}
&G_{FI,\lambda} = - a_0 \int_0^\infty R_\perp dR_\perp \int_{-\infty}^\infty dZ
					\nonumber\\
&\hskip 1 em	\times	\mathcal Y_F^*(R_\perp,Z)  \,
	J_{m'_R-m_R-\lambda}(\kappa R_\perp) \, e^{ik_z Z} \,
	\partial_{R\lambda} \mathcal Y_I(R_\perp,Z)	\,.
\end{align}

In general, unless the CM state is extremely localized, so that the derivatives of the $\mathcal Y_I$ are large, the contributions coming from the derivative acting on the CM coordinate, $\mathcal M_{e,R}$, are small compared to the term where the derivative acts on the relative coordinate $\mathcal M_{e,rel}$. 

The cases where the proton absorbs the photon can be similarly written down, and will be numerically small because of the mass factors.  

All the photoproduction amplitudes display a $\delta$-function that explicitly shows conservation of angular momentum projection along the $z$-direction. This assures us that the initial total helicity either goes into excitation of the atomic state, or into revolution of the entire atom about the origin defined by the twisted photon's symmetry axis, or into the sum of the two.  The relative amount of angular momentum passed to the CM motion of the whole atom compared to the angular momentum projection of the excited electronic state depends on the impact parameter $b$, for cases where the CM wave function is sharply peaked about $R_\perp = b$ in Eq~(\ref{eq:GFI3}) or $R=b$ and $\cos\theta_R\approx 0$ in Eq.~(\ref{eq:twiddle}).


\subsection{$1S\to 2P$ atomic photoexcitation}


For ordinary plane wave photons with helicity $\Lambda$, a photoexcited $2P$ state necessarily has $m_f = \Lambda$, if we start from the ground state.   For a twisted photon initial state, other final quantum numbers (QN) are possible.  In particular, for a twisted photon with $m_\gamma = 0$,  the zero angular momentum projection final state may be produced with sufficient amplitude to be observable.

We will take the approximation that amplitude $\mathcal M_{e,rel}$ is most of the amplitude and use the notation $\mathcal M^{(m_\gamma)}_{n_f l_f m_f}$ to keep track of the transitions from and to different QN states.  We calculate amplitudes from Eq.~(\ref{eq:localized}) or from the localized version of the definite $L_Z$ states mentioned after Eq.~(\ref{eq:twiddle}).

We take $\Lambda = 1$ for definiteness.  For $m_\gamma = 0$, examining Eq.~(\ref{eq:localized}), we can produce only the $m_f = 0$ final state if the target atom is on the photon axis, and progressively produce more of the $m_f = 1$ final state when the target atom is off the axis.   The largest of the atomic factors for the $m_f = 0$ is the one labeled $g_{2100}$, and similarly for $m_f=1$ the largest atomic factor is $g_{2111}$.  In the dipole approximation, $k \ll a_0$, these are the same, either by the Wigner-Eckart theorem or by direct examination,
\begin{align}
g_{21\lambda\lambda} &=  \frac{1}{2\pi} \int_0^\infty r^2 dr \, R_{21}(r)  R_{10}(r)
= \frac{1}{\pi} \left( \frac{2}{3} \right)^{7/2} 	.
\end{align}
Hence the main difference between the peak values of the amplitudes for $m_f = 0$ and $m_f = 1$ final states produced from the special $m_\gamma = 0$ twisted photon comes because the pitch angle $\theta_k$ factors are different in the different terms of the overall amplitude, Eq.~(\ref{eq:localized}).

\begin{figure}[htbp]
\begin{center}

\includegraphics[width = 78 mm]{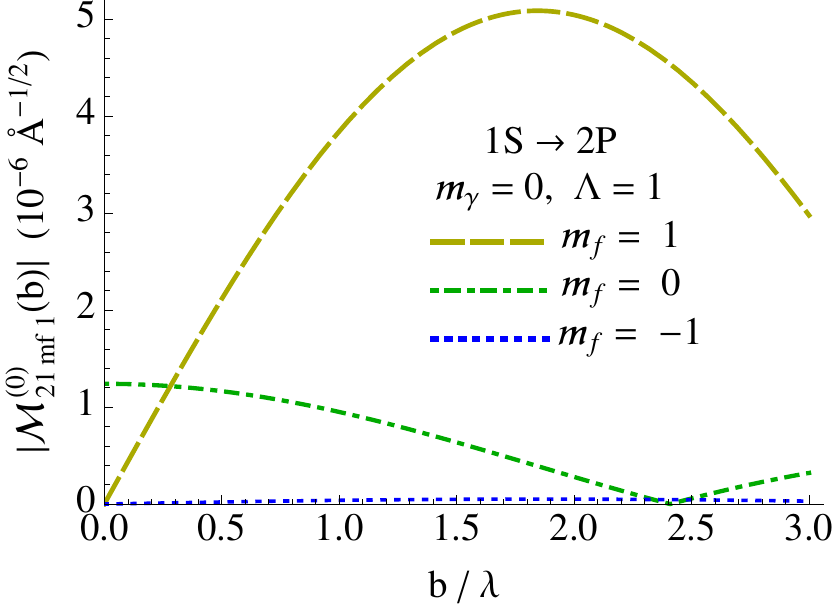}

\caption{$1S\to 2P$ photexcitation amplitudes for $m_\gamma = 0$ twisted photons, with pitch angle $\theta_k=0.2$ radians.}
\label{fig:amplitudes0}
\end{center}
\end{figure}

Fig.~\ref{fig:amplitudes0} plots the absolute values of the amplitudes for the $1S\to 2P$ transitions for $m_\gamma = 0$ and the different $m_f$ as a function of the transverse distance of the target from the photon axis (measured in photon wavelengths), for a pitch angle $\theta_k = 0.2$ radians.


\begin{figure}[htbp]
\begin{center}

\includegraphics[width = 78 mm]{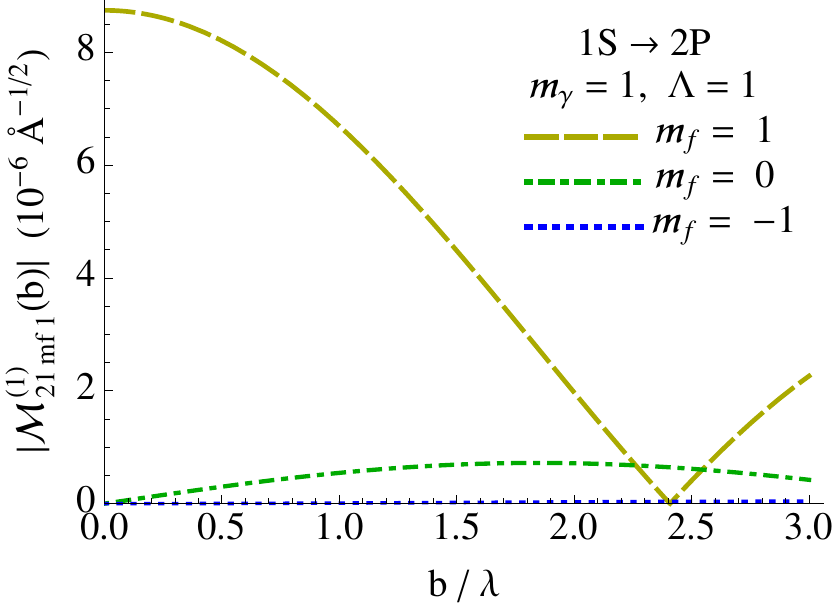}

\caption{$1S\to 2P$ photexcitation amplitudes for $m_\gamma = 1$ twisted photons, with pitch angle $\theta_k=0.2$ radians.}
\label{fig:amplitudes1}
\end{center}
\end{figure}

For comparison, the corresponding plot is shown in FIg.~\ref{fig:amplitudes1} for the $m_\gamma = 1$ twisted photon, which has essentially the same peak amplitudes as the plane wave case.  The magnitudes are not significantly larger than the peaks in the $m_\gamma = 0$ case.  


\section{Forces Induced by Twisted Photons
}			\label{sec:tractor}


There has been recent discussion regarding tractor beams, meaning propagating beams which attract particles in their path.  In is known, experimentally as well as theoretically, that one can attract small particles using gradient forces~\cite{2011PhRvL.107t3602S,2012PhRvL.109p3903R,1996OptCo.124..529H,1986OptL...11..288A}.  This is done using interfering beams to produce a varying electric field magnitude along the beam axis, and then to move the equilibrium points of the resulting force by changing the relative phase of the beams.

Contention has surrounded the possibility of non-gradient tractor beams.  It has been suggested that there could exist propagating field configurations which would pull small particles all the way back to the source, with no field gradient along the beam axis and without time-dependent manipulation of the emitted beams~\cite{Novitsky_2007,Chen2011,Saenz_2011,Novitsky_2011,zhang2011pre,Bekshaev:2011,dogariu2013natpho}.

We shall consider how this discussion relates to the Bessel twisted photons we have discussed here.  The Bessel beam Poynting vector, $\vec S = \vec E \times \vec B$, in components is
\begin{align}
\label{eq:poynting}
S_\rho &= 0	\,,	\nonumber\\
S_\phi &= \frac{\kappa \omega^2}{4\pi} \sin\theta_k \, J_{m_\gamma}(\kappa\rho)
		\nonumber\\
& \quad \times	\left( \cos^2\frac{\theta_k}{2} J_{m_\gamma-\Lambda}(\kappa\rho)
		+ \sin^2\frac{\theta_k}{2} J_{m_\gamma+\Lambda}(\kappa\rho)	\right)  \,,	
					\nonumber\\
S_z &= \frac{\kappa \omega^2}{4\pi}
	\left( \cos^4\frac{\theta_k}{2} J^2_{m_\gamma-\Lambda}(\kappa\rho)
		- \sin^4\frac{\theta_k}{2} J^2_{m_\gamma+\Lambda}(\kappa\rho)	\right)	\,,
\end{align}
using the normalization of Ref.~\cite{Jentschura:2010ap,Jentschura:2011ih}.  The Poynting vector for these states is naturally time independent.  Note the presence of a transverse component above, as well as the presence of a longitudinal component~\cite{Barnett:1994OptCo.110..670B} in the electric field and in the potential, Eq.~(\ref{eq:pot}).  

There are regions where the Poynting vector points back towards the source.  For example, consider the Poynting vector on-axis for $m_\gamma = -\Lambda$.  In simple cases, the radiation pressure is given by the Poynting vector, but this is not true in general.  Here follows a brief discussion, followed in turn by consideration of the forces on small particles by a Bessel beam.

Electromagnetic forces are given most directly by the Lorentz force law, and for small, neutral, polarizable particles this may be expressed as
\begin{align}
\vec F = ( \vec p \cdot \vec \nabla) \vec E + \dot { \vec p } \times \vec B		\,,
\end{align}
where $\vec p$ is the electric dipole moment of the particle.  ``Small'' means size scales small compared to the wavelength of the electromagnetic radiation.   If all time dependences are monochromatic, $\vec p = \Re ( \vec p_0 \, e^{-i\omega t})$, $\vec E = \Re (\vec E_0 \, e^{-i\omega t})$, \textit{etc.}, and if there is a linear relation  $\vec p_0 = \alpha \vec E_0$, then the time averaged force $\langle \vec F \rangle$ is given by~\cite{Chaumet_2000,Ruffner_2013}
\begin{align}
\label{eq:ruffgrier}
\langle \vec F \rangle = \frac{1}{2} \Re \left( \alpha \vec E_0 \cdot (\vec\nabla) \vec E_0^* 
	\right)	\,.
\end{align}
Here, 
$[ \vec E_0 \cdot (\vec\nabla) \vec E_0^* ]_i = \sum_{j=1}^3 E_{0j} \nabla_i  E_{0j}^* $.  The polarazability may be complex, reflecting a phase difference between $\vec P$ and $\vec E$, so that
\begin{align}
\langle \vec F \rangle = \frac{1}{4} \left( \Re \alpha \right)  \vec\nabla | E_0 |^2 
	- \frac{1}{2} \left( \Im\alpha \right) \Im \left( \vec E_0 \cdot (\vec\nabla) E_0^* \right)
	\,;
\end{align}
the first term is the gradient force.  The second term, when $\vec\nabla\cdot\vec E = 0$ is allowed, can be further manipulated to give~\cite{Albaladejo_2009,Marques_2013}
\begin{align}
\label{eq:alba}
\langle \vec F \rangle &= \frac{1}{4} \left( \Re \alpha \right)  \vec\nabla | \vec E_0 |^2 
	+ \sigma \langle \vec S \rangle 
	+ \frac{\sigma}{4i\omega} \vec\nabla \times \left(   \vec E_0 \times \vec E_0^* \right)
	\,,
\end{align}
where $\sigma = \omega (\Im \alpha)$, and $\vec E_0 \times \vec E_0^*$ is related to the spin angular momentum of the electromagnetic wave (see, for example, Eq.~(14.22) of~\cite{Bjorken-Drell2}). The last term is the spin-curl term, and is zero for simple plane wave states.

The forms Eq.~(\ref{eq:ruffgrier}) and Eq.~(\ref{eq:alba}) are generally equivalent.  One has the virtue of showing how the radiation pressure comes directly from the Poynting vector in some cases.  The other is simpler to write down, and for the Bessel beams gives the result for the force more directly.

Regarding the force on a small particle for which we have monochromaticity and $p_0 = \alpha E_0$, specifically for the case where the particle is on the axis of a $m_\gamma = -\Lambda$ Bessel beam, the Poynting vector points opposite to the propagation direction. This would give a tractor beam if the sometimes accepted direct connection between the Poynting vector and the radiation force were correct.  However, using Eq.~(\ref{eq:ruffgrier}), one sees the longitudinal force involves directly the derivative $\nabla_z$.  In the Bessel beams, the only $z$-dependence is in the exponential $\exp{i k_z z}$,  and 
\begin{align}
\langle F_z \rangle = \sigma | \vec E_0 |^2 \cos\theta_k	\,,
\end{align}
which is always positive for a forward propagating Bessel beam if $(\Im\alpha)$ is positive, which it is whether it comes from radiation reaction forces or from drag forces within the target particle.

Using Eq.~(\ref{eq:alba}) leads to the same result.  The Poynting term is just
\begin{align}
\langle F_z \rangle_{\rm Poynting} = - \sigma | \vec E_0 |^2	\,,
\end{align}
using the information that the Poynting vector for the $m_\gamma = -\Lambda$ beam on axis has the same magnitude as $| \vec E_0 |^2$ but a negative sign.  However, the spin-curl force is not zero, and gives 
\begin{align}
\langle F_z \rangle_{\rm sc} = 2 \sigma | \vec E_0 |^2 \cos^2 \frac{\theta_k}{2}	\,.
\end{align}
The sum of the spin-curl force and the Poynting term gives the same result as found farther above.

We will also make some remarks regarding forces on totally absorptive particles.  Monochromaticity does not apply.  At equilibrium, totally absorptive bodies will balance absorbed energy with (non-monochromatic) black body radiation.  However, one can proceed by writing the Lorentz force law in terms of the stress tensor.  If the Poynting vector is bounded, the time average force is given by a areal integral over a closed surface surrounding the particle in question,
\begin{align}
\langle \vec F \rangle = \oint da \ \hat n \cdot \langle \tensor T \rangle		\,,
\end{align}
where $\hat n$ is the outward normal.  The stress tensor is
\begin{align}
T_{ij} = E_i E_j - \frac{1}{2} \delta_{ij} \vec E^2 + B_i B_j - \frac{1}{2} \delta_{ij} \vec B^2 \,.
\end{align}

For a simple example, consider a small flat thin object, on-axis, facing the beam.  Integrate over a thin surface barely outside the object.  If the object is totally absorptive, the fields just before the object will be those of the unpolarized Bessel wave, and they will be zero just behind the object.  The only relevant normal is $-\hat z$ on the front surface, and 
\begin{align}
\langle F_z \rangle = - \int_{\rm front\ surface} da  \, \langle T_{zz} \rangle	\,.
\end{align}
If the object is small and of cross section area $\sigma$, this gives 
\begin{align}
\langle F_z \rangle = \sigma | \vec E_0 |^2		\,,
\end{align}
for the $m_\gamma = - \Lambda$ example.


\section{Summary and Discussion}			\label{sec:disc}


We have discussed twisted photons, or photons whose total angular momentum projected along the direction of wavefront propagation can be very large.  These photons, unlike plane wave photons, have with a Poynting vector that is not zero in the transverse plane and swirls about a symmetry axis.  We have, in particular, focused on two applications, photoexcitation of hydrogen-like atomic states and situations where the Poynting vector along the propagation direction is reversed. 

Regarding atomic photoexcitation, twisted photons offers unique opportunities to produce final states with arbitrary angular momentum projections, using the direction of motion of the incoming wavefront as a quantization axis.  We have extended our previous photoexcitation analysis~\cite{Afanasev:2013kaa} to show how in our formalism photon angular momentum divides conservatively between rotating the CM state and angular momentum of the excited state (see also~\cite{Jauregui:2004}).  For a specific example, instead of a very large orbital angular momentum, we considered projected angular momentum $m_\gamma = 0$, a value also not possible with plane wave photons.  This allows leading order, in the atomic matrix elements, $1S\to 2P$ transitions with final magnetic quantum number $m_f = 0$.  The rates are suppressed by factors dependent upon the pitch angle (the angle from the twisted photon propagation direction to its component plane wave states), but could be large enough to notice if the pitch angle is not small.

We observe that there are regions where the Poynting vector points back toward the light source.  Any region where the first Bessel function in the equation for $S_z$ (Eq.~(\ref{eq:poynting})) is zero or near zero will serve as an example.  The negative sign term is suppressed by several powers of the pitch angle, so will not be found with a paraxial approximation, nor will it be found if polarization is neglected.  A particular example is the near-axis case with $m_\gamma = 1$ and component plane wave photon helicity $\Lambda = -1$ for which the negative sign Poynting component is maximized.  

However, in a complete analysis, as presented here, the total radiation pressure remains positive.  This has been noted previously in the 
literature~\cite{Albaladejo_2009,Ruffner_2013},  but not verified explicitly for the circularly polarized Bessel beams with orbital angular momentum considered in this paper.


\appendix

\section{Comparison of twisted photon expansions}			\label{sec:comparison}

For a given \{$\kappa, m_\gamma, k_z$\}, we have a pair of independent states with $\Lambda = \pm 1$.  For the same \{$\kappa, m_\gamma, k_z$\}, 
J\'auregui~\cite{Jauregui:2004} also has a pair of states, labeled TE and TM.  The two pairs of states are equivalent, as we shall show.  

Using cylindrical coordinates 
$\vec r = (r_\perp,\phi_r,z)$ and
\begin{align}
\psi_m = \psi_m(r_\perp,\phi_r,\kappa) 
	= J_m(\kappa r_\perp) e^{im_\gamma \phi_r}	\,,
\end{align}
the TE and TM modes from~\cite{Jauregui:2004} are
\begin{align}
\vec A^{(\rm{TE})}_{\kappa m_\gamma k_z}(t,\vec r) &=
	\frac{ i E_0 }{k_z\sqrt{2}}  e^{i(k_z z - \omega t)}
	\left(  \psi_{m_\gamma -1} \hat \eta_-  - \psi_{m_\gamma +1} \hat \eta_+  \right)	\!,
			\nonumber\\
\vec A^{(\rm{TM})}_{\kappa m_\gamma k_z}(t,\vec r) &=
	-\frac{ E_0 }{\omega \sqrt{2}}  e^{i(k_z z - \omega t)}	\nonumber\\
&\hskip -2.5 em  \times
	\left( \psi_{m_\gamma -1} \hat \eta_- + \psi_{m_\gamma +1} \hat \eta_+ 
	+ i \sqrt{2} \tan\theta_k \, \psi_{m_\gamma}  \hat\eta_0  \right)	.
\end{align}

With $\Lambda = \pm 1$, the states we use are
\begin{align}
\vec A_{\kappa m_\gamma k_z 1} &= \frac{1}{i} \sqrt{\frac{\kappa}{2\pi}}
	e^{i(k_z z - \omega t)}	
	\bigg\{ \psi_{m_\gamma - 1} \cos^2\frac{\theta_k}{2} \,\hat\eta_+
				\nonumber\\
&\hskip 3 em			- \psi_{m_\gamma + 1} \sin^2\frac{\theta_k}{2} \,\hat\eta_-
			+ \frac{i}{\sqrt{2}} \psi_{m_\gamma} \,\hat\eta_0
	\bigg\}	,
				\nonumber\\
\vec A_{\kappa m_\gamma k_z, -1} &= \frac{1}{i} \sqrt{\frac{\kappa}{2\pi}}
	e^{i(k_z z - \omega t)}	
	\bigg\{ \psi_{m_\gamma - 1} \sin^2\frac{\theta_k}{2} \,\hat\eta_+
				\nonumber\\
&\hskip 3 em			- \psi_{m_\gamma + 1} \cos^2\frac{\theta_k}{2} \,\hat\eta_-
			- \frac{i}{\sqrt{2}} \psi_{m_\gamma} \,\hat\eta_0
	\bigg\}	.
\end{align}
Hence,
\begin{align}
\vec A^{(\rm{TE})}_{\kappa m_\gamma k_z} &= \frac{ E_0}{k_z}
	\sqrt{\frac{\pi}{\kappa}}	\left( \vec A_{\kappa m_\gamma k_z 1}
		+ \vec A_{\kappa m_\gamma k_z, -1}	\right)	\,,
			\nonumber\\
\vec A^{(\rm{TM})}_{\kappa m_\gamma k_z} &=  - \frac{iE_0}{k_z}
	\sqrt{\frac{\pi}{\kappa}}	\left( \vec A_{\kappa m_\gamma k_z 1}
		- \vec A_{\kappa m_\gamma k_z, -1}	\right)	\,.
\end{align}
The relations are clearly invertible, so the expansions are equivalent.


\begin{acknowledgments}

CEC thanks the National Science Foundation for support under Grant PHY-1205905 and thanks the IIT-Bombay for hospitality during part of the time this work was underway. Work of AA was supported by The George Washington University.  We thank Konstantin Bliokh for useful comments and the participants and organizers of the ``Structured Light in Structured Media: from Classical to Quantum Optics Incubator'' meeting for stimulating discussions.

\end{acknowledgments}


\bibliography{TwistedPhoton2}

\end{document}